# Photonic Altermagnets: Magnetic Symmetries in Photonic Structures

*Andrew S. Kim[1,2,3*], Youqiang Huang[4], Zhipei Sun[4], Q-Han Park[5], and Hyunyong Choi[2,3*]*

[1] Research Institute of Basic Science, Seoul National University, Seoul, 08826, Republic of Korea

[2] Department of Physics and Astronomy, Seoul National University, Seoul, 08826, Republic of Korea

[3] Institute of Applied Physics, Seoul National University, Seoul, 08826, Republic of Korea

[4] Department of Electronics and Nanoengineering, Aalto University, Espoo 02150, Finland

[5] Department of Physics, Korea University, Seoul, 02841, Republic of Korea

*E-mail: a.s.kim@snu.ac.kr, hy.choi@snu.ac.kr

**Abstract**

The unique physical properties of altermagnets, when transplanted to photonic systems, are anticipated to offer a new degree of freedom for engineering electromagnetic waves. Here, we show that a photonic analogue of altermagnetism can be mimicked in photonic crystals, where engineered photonic crystals can host spin space group symmetries. Our approach allows for the creation of spin-split bands and the corresponding transport properties provide an effective platform for circularly polarized light isolation without the need of geometrodynamic spin-orbit interaction. Beyond the concurrent solid-state materials, we anticipate our work to offer photonic crystals as a versatile platform to test the spin-split band properties and inspire optical designs for photospintronic applications.

**Main**

The conventional classification of collinear magnetic orders is now challenged by the recent discovery of altermagnetism, revealing that certain types of collinearly-ordered antiferromagnets display broken Kramers degeneracy despite vanishing net magnetic moments [1-5]. This new phase features two spin sublattices that cannot be transposed solely through translation or inversion but also requires rotational transformation, asserting the major contribution of non-magnetic ions to time-reversal symmetry (TRS) breaking [6,7]. Altermagnets combine the advantages of both ferromagnets and antiferromagnets [8-11], presenting significant promises for spintronic applications [12,13]. However, studying their physical properties and developing spintronic devices remain experimentally challenging, where only recently have altermagnetic spin-split bands been experimentally confirmed [14-18].

Photonics has long stood as a unique platform for exploring concepts derived from solid-state physics. The similarity in the mathematical forms between Maxwell's equations and Schrodinger's equation has inspired the engineering of artificial crystal lattices for photons, namely photonic crystals (PhCs) [19-21]. Importing solid-state phenomena to the photonic domain including bandgaps [22-26], Weyl points [27], topological edge states [28,29], and spin-valley locking [30] have enabled photonic devices with complex controllability of light. Following this line of research, PhCs naturally emerge as a promising platform for demonstrating altermagnetic spin-split bands. Similar to electronic systems, momentum-dependent chiroptical response can arise in certain types of achiral structures. Here, chiroptical response refers to the helicity-dependent interactions between circularly polarized light and chiral media—materials with broken mirror symmetry. Termed "pseudochirality", this property has been extensively studied owing to its facile tunability of chiroptical response through azimuthal rotation of obliquely incident light [31,32]. Considering the analogy between electron spin and circularly polarized light, it seems evident that there is an intricate connection between altermagnetic symmetries and pseudochirality.

In this work, we establish a theoretical framework which allows altermagnetic spin-split bands to be rendered in PhCs. Key ingredients of this framework include the construction of a symmetry operator connecting orthogonal polarization states into Kramers doublets [33], and the introduction of chirality as a means to lift the symmetry-protected Kramers degeneracy of photon spin pairs. Owing to their parallelism to spin space group symmetries, these building blocks enable photonic systems to exhibit spin-split bands analogous to those found in the entire family of collinearly-ordered magnetic phases. Accordingly, the *d*-wave symmetric altermagnetic spin-split bands and their subsequent spin-dependent transport can be reproduced solely through the replication of the spin space group symmetry arguments. Our framework suggests the potential of photonic platforms as experimental testbeds for the emerging field of altermagnetism and ultimately magnetic spin-split band structures toward photospintronic applications [34,35].

Magnetic phases can be classified by inspecting the symmetry properties of magnetic materials in real and reciprocal spaces [6], where similar arguments are applicable to photonics based on a few correspondences. First, as presented in Fig. 1a, circularly polarized light (or photon helicity states) possesses spin-angular momentum (SAM) and therefore is often associated with electron spin. Just as spin-split bands arise from non-zero net magnetic moments in magnetic materials, chiroptic media exhibit helicity-dependent optical dispersions [36]. Thus, as shown in Fig. 1b, objects with opposite chirality make a good analogy between magnetic atoms with opposite magnetic moments. Finally, we propose that the time-reversal operation in electronic systems corresponds to the chirality switching operation in photonic systems, which will be further justified. Derived from this notion, the periodic arrangement of chiroptic objects with uniform (alternating) handedness serve as photonic analogues of ferromagnets (antiferromagnets).

To further justify our analogy, we begin by examining how photon helicity pairs can form Kramers doublets. In contrast with electrons, photons behave differently under time-reversal operation due to differences in their spin numbers, where helicity is preserved rather than being reversed. Thus, photons with opposite helicities are not a Kramers doublet connected through standard TRS. Nevertheless, it is evident that Kramers degeneracy exists in photonic systems once we realize that arbitrarily polarized photons always have orthogonal counterparts with the same energy in free space. This arises from the invariance of the Maxwell's equations to parity ($\mathcal{P}$), bosonic time-reversal ($\mathcal{T}_b$), and duality exchange ($\mathcal{D}$) operations [37,38] in free space. While none of these symmetry operations alone guarantee the transformation of arbitrary polarization states to the orthogonal counterparts, their combination into a pseudo-time-reversal (pTR) symmetry operator $\mathcal{T}_p = \mathcal{P}\mathcal{T}_b\mathcal{D}$ enables such interconversion. Intuitively, the parity operation converts the helicity of photons and the reversal of photon momentum is compensated through the bosonic time-reversal operation. Finally, linear polarization states can be converted to their orthogonal states through the duality exchange operation. Beyond interconverting the Kramers pairs, $\mathcal{T}_p$ also affects the handedness of chiral materials. The inclusion of the parity operator ($\mathcal{P}$) within $\mathcal{T}_p$ allows it to flip both the helicity of photons and the handedness of chiral objects, akin to how fermionic TR flips electron spins and magnetic moments. This suggests that the chirality-switching operator is none other than $\mathcal{T}_p$, signifying its role as a photonic analog of fermionic TRS.

Based on such correspondence, we construct a PhC that follows altermagnetic symmetries. The unique spatial arrangement of non-magnetic atoms constitutes two anisotropic spin sublattices in altermagnets. Despite having the same magnetic arrangement as antiferromagnets, rotation operation is required to superimpose the two spin sublattices which breaks the TRS. Applying this principle to photonics, we first consider a two-dimensional (2D) PhC consisting of infinitely long cylinders with chiroptic responses. Initially, the cylindrical chiral objects are arranged to form an array of periodically alternating handedness, thereby constructing a photonic antiferromagnet. Now, in photonic structures, the inclusion of non-magnetic atoms can be omitted by introducing anisotropy to the chiral object itself. As illustrated in the left panel of Fig. 1c, we transform the chiral objects into elliptic cylinders and perform an additional $C_4$ rotation for chiral objects of certain handedness. With such an arrangement, we see that the pTR pair of the constructed PhC can be obtained through simple $C_4$ rotation. This precisely follows the spin space group symmetry argument of *d*-wave altermagnets [7]. In this configuration, it is the anisotropic spatial distribution of the chirality parameter that leads to the orientation-dependent handedness. The altermagnetic PhC can be approximated as an effectively homogeneous media when the lattice constant of the PhC unit cell is smaller than the photon wavelength. In this limit, group theory [39,40] suggests that altermagnetic PhCs are equivalent to effectively homogeneous pseudochiral media [41] (see Supplemental Material, Sec. I). A typical dispersion relation (Fig. 1c, middle) and the isofrequency contour (Fig. 1c, right) of a pseudochiral media exhibit notable similarities to the characteristic spin-split bands and Fermi surfaces of altermagnets, hinting at the validity of our inspection. For detailed theoretical proofs of altermagnetic Kramers degeneracy lifting in both effective material and photonic crystal limits, refer to Supplemental Material (Sec. II–III).

We further note that the physical traits of altermagnetism are also reflected in the mathematical properties of pseudochiral bianisotropic tensors. In the effective medium limit, the electromagnetic response of both chiral and pseudochiral media can be characterized by additionally including bianisotropic tensors as effective parameters in the constitutive relations of electromagnetic fields. These tensors quantify the induced electric (magnetic) polarization in response to an external magnetic (electric) field. As shown in Supplemental Material Sec. III, the signs of both chiral and pseudochiral bianisotropic tensors reverse under the $\mathcal{T}_p$ transformation, indicating that both chirality and pseudochirality break the pTR symmetry. Additionally, pseudochiral bianisotropic tensors consist only of symmetric off-diagonal components, ensuring the trace of the bianisotropic tensor to be zero. Since the trace represents net chirality, this confirms that pseudochirality shares a key feature of altermagnetism, where Kramers degeneracy is lifted despite the absence of net magnetization.

Now, we investigate the helicity-split band structure of a model photonic altermagnet in-depth via numerical analysis (see Supplemental Material, Sec. IV), where energy eigenvalues for different propagation vectors are calculated. We follow the configuration illustrated in Fig. 1c, which is a 2D PhC consisting of isotropic and homogeneous materials. The material parameters are set to ensure the duality symmetry, which holds as long as the ratio between permittivity and permeability tensors remains a constant scalar value $\eta_r^2$ throughout the entire domain [44,45]. Reflecting the symmetry constraints, we fix $\eta_r$ to 1 throughout the entire system. The relative permittivity ($\varepsilon_r$) and permeability ($\mu_r$) of the achiral embedding matrix are both set to 1, respectively, and the relative permittivity ($\varepsilon_r$) and permeability ($\mu_r$) of the chiral cylinders are both set to 2. The magnitude of the chirality parameter is set to 1.5 with a positive (negative) sign for the right (left) handed chiral cylinder. The geometry is defined by setting the ratio between the lattice constant ($a_0$) and the diameter of the cylinder (D) as $D/a_0 = \frac{3\sqrt{2}}{10}$. At this stage, spatially arranging chiral objects according to their handedness forms photonic ferromagnets or antiferromagnets. Indeed, the calculations on their photonic helicity-split band reveal the characteristic spin-split band behavior of their solid-state counterparts, as presented in the Supplemental Material (Sec. V).

We then transform the chiral cylinders into elliptic cylinders described by an asymmetry parameter α with a value of 1.3. Here we assume a standard unit cell, where the rhombic unit cell shown in Fig. 1c is transformed to an ordinary square unit cell through 45° rotation. The numerical results are shown in Fig. 2a, where normalized *z*-component Riemann-Silberstein (RS) vectors are presented. Considering the RS vectors as the photon wavefunctions with well-defined helicity [42,43], we see that photons are localized near the chiral objects of specific handedness depending on the helicity eigenstates. Such a characteristic behavior and the resulting spatial profile resemble the spin-density isosurfaces of *d*-wave altermagnets, where electrons of distinct spin are confined within one of the two spin sublattices [44]. The photonic band diagram in Fig. 2b further reveals the effect of the pTR symmetry breaking. Similar to the altermagnets in solids, we see that the energy bands are helicity-polarized and exhibit the altermagnetic lifting of the Kramers degeneracy. Hence the symmetry also impels helicity-degeneracy within the $\Gamma M(M')$ interval. The symmetry-protected degeneracy comes in a form of four nodal points in IFCs. Figure 2c shows exemplary IFC maps, sampled at two different normalized photon frequencies. The IFC of the lower frequency (Fig. 2c, bottom) resembles the IFCs of the pseudochiral metamaterials [45], however with a less pronounced helicity-splitting effect. As seen in Fig. 2c, the helicity-dependent splitting of IFCs becomes prominent for the higher energy bands. Note that the enhanced band splitting at higher energies make spin-split bands observable even for small chirality parameter values (see Supplemental Material, Sec. VI).

The helicity-split bands observed in the proposed altermagnetic PhC host unique helicity-dependent wave propagation dynamics, hinted by their solid-state counterpart. Taking a *d*-wave symmetric altermagnet for example, spin-dependent transport dynamics such as the anomalous Hall effect (AHE) and the altermagnetic spin-splitting effect (ASSE) exist. Contrary to AHE which requires net non-zero Berry curvature and thus the presence of spin-orbit coupling (SOC) [1], ASSE doesn't require SOC. The nonrelativistic nature allows ASSE to exhibit superior spin conductivity over AHE [2], where the large propagation angle between electrons with opposite spin directions is simply dictated by the anisotropic spin-split bands. Illustrated in Fig. 3a (left), we show that ASSE can be repeated in PhCs as a photonic spin-splitter effect (PSSE). Similar to electron transport, when the propagation direction of an incident linearly polarized light points towards the symmetry-protected nodal points (i.e., the ΓX and ΓY directions), the linearly polarized light splits into two beam paths with opposite helicities. As described in Fig. 3a (middle), the photon group velocity within the PhC is directed normal to the helicity-dependent IFCs, resulting in the helicity-dependent photon deflection. Numerical calculations validate the existence of the PSSE. As depicted in Fig. 3a (right), given a linearly-polarized Gaussian beam with the frequency $\omega a_0/2\pi c = 0.62$ incident on the photonic altermagnet, we verify the helicity-multiplexed angle of refraction where the handedness of the Gaussian beam is indicated through color-coding. We emphasize that the Berry curvature does not exist in the altermagnetic PhC, indicating that the 'relativistic' effects are indeed excluded in our setup (see Supplemental Material, Sec. VII for numerical calculations). We therefore infer that PSSE is simply the result of broken Kramers degeneracy instead of the manifestation of strong spin-orbit interaction. This makes PhCs an appealing platform, since the existence of Berry curvature in solids complicates the separation of contributions from AHE and ASSE in spin current measurements.

In solid-state altermagnets with similar spin-space group symmetries, applying an electric field and, therefore, inducing electric currents along the $\Gamma M(M')$ direction will generate spin-polarized currents. Although the underlying mechanism is different, similar outcomes in altermagnetic PhCs appear yet in a pronounced manner (Fig. 3b, left). Inspecting the sampled IFC (Fig. 3b, middle), within certain ranges of propagation direction, we see that photonic Bloch states exist only for specific helicity states. Under these conditions, the altermagnetic PhC transmits only one of the two helicity states and completely reflects the other, rendering it as a perfect photonic spin filter. The photonic spin filtering (PSF) is confirmed through full-wave simulations under an incident vertically polarized Gaussian beam, as shown in Fig. 3b (right). Intriguingly, the color textures surrounding the incident Gaussian beam in Fig. 3b (right) show that the helicity of the reflected light is opposite to the transmitted light (see Supplemental Material, Sec. VIII). This implies that the reflected beam retains the same handedness as the rejected portion of the incident light. Therefore, the altermagnetic PhC can function not only as lossless spin filters but also as helicity-preserving mirrors. Indeed, the helicity-selectivity of the altermagnetic PhC can be switched by rotating the light propagation direction due to its inherent anisotropy, which is also observed in the simulation results.

In summary, we show that the correspondence between chiroptics and magnetism originates from a shared foundation in their symmetry principles, thereby reinterpreting pseudochirality through the perspective of spin space group symmetries. Spin-split bands and their resulting spin-dependent transport properties (PSSE and PSF) in altermagnetic PhCs are unprecedent. Particularly, the PSSE has been identified to be distinct from conventional geometric-phase based helicity-splitting methods. Moreover, the uniaxial nature of altermagnetic PhCs makes PSSE an in-plane propagation mode effect, thereby presenting promising opportunities for integrated spin-optic circuitry. Notably, with judicious design, momentum-dependent optical activity can also be achieved, having potential applications for optical information processing (see

Supplemental Material, Sec. IX). Considering that studies on altermagnetism yet remain in their early stages, the following perspectives are outlooked. First, the controllability of photonic crystal elements provides opportunities for circumventing the current experimental hurdles of domain control in solid-state altermagnets. Second, the ability to trace electromagnetic fields in PhCs allows the investigation of spin-split photon trajectory followed by PSSE, which is currently beyond experimental reach for electrons in solids. Third, translating the unique spin-space-group symmetry into distant fermionic (e.g. *d*-wave superconductivity [6,7,46]) or bosonic (e.g. photonics) domains may challenge our understanding of the universality of such symmetry types. Lastly, our framework further extends towards the possibility of designing unconventional even-parity wave pseudochirality in photonics inspired by altermagnetic materials [7,14,17], which hasn't been contemplated so far.

ASSOCIATED CONTENTS

**Supplemental Material**

Supplemental Figures and Supplemental Notes.

AUTHOR INFORMATION

Corresponding Author

E-mail: a.s.kim@snu.ac.kr, hy.choi@snu.ac.kr.

**Notes**

The authors declare no competing financial interest.

**Data Availability**

All study data are available in the article main text or in the Supplemental Material.

**Acknowledgements**

The authors thank Dr. Chris Fietz for his kind consultation on the numerical simulations for the photonic band structures. This research was supported by the National Research Foundation of Korea (NRF) through the government of Korea (Grant No. 2021R1A2C3005905, RS-2024-00413957, RS-2024-00466612, RS-2024-00487645), Scalable Quantum Computer Technology Platform Center (Grant No. 2019R1A5A1027055), the Institute for Basic Science (IBS) in Korea (Grant No.IBS-R034-D1), Global Research Development Center (GRDC) Cooperative Hub Program through the National Research Foundation of Korea (NRF) funded by the Ministry of Science and ICT (MSIT) (Grant No. RS-2023-00258359), and the core center program (2021R1A6C101B418) by the Ministry of Education.

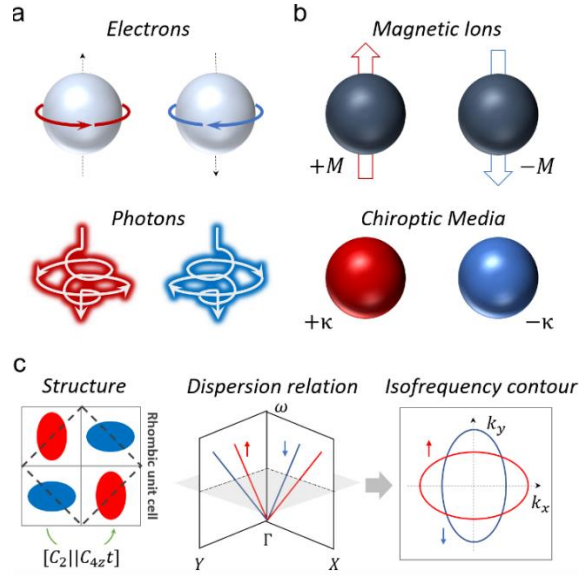

**Fig. 1.** Illustration of the corresponding entities between condensed matter and photonic systems. (a) The spin quantum numbers relate the electron spins to the photon helicity, where positive and negative spin numbers are color-coded as red and blue. (b) Subsequently, according to their spin/helicity-dependent responses, magnetic ions with magnetic moments (M) pointing up (red) and downwards (blue) in the out-of-plane direction correspond to chiroptic media with right (red) and left (blue) chirality (κ). (c) Layout of a photonic crystal altermagnet assuming the analogy between magnetic moment reversal and chirality switching (left). The rhombic unit cell includes two alternating chiral objects and the equivalent of spin space group operations—magnetic moment reversal ($C_2$), 90° spatial rotation about the z-axis ($C_{4z}$), and spatial translation ($t$)—restoring the initial states of spin sublattices in solid-state altermagnets can be applied. A qualitative depiction of the optical dispersion (middle) and the isofrequency contour (right) can be analytically obtained through Maxwell's equations with pseudochiral material tensors.

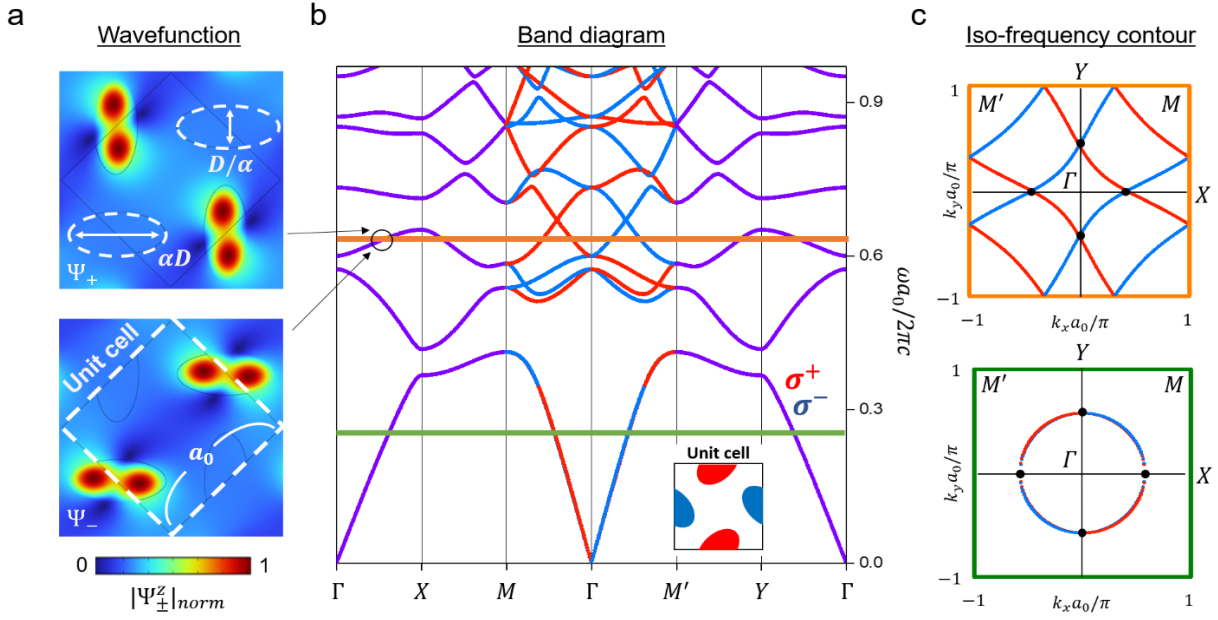

**Fig. 2.** Helicity-polarized band properties of photonic altermagnets. (a) Normalized field distribution maps for right (top) and left (bottom) handed photons sampled at the normalized frequency $\omega a_0/2\pi c = 0.62$ within the $\Gamma X$ interval. The field distribution resembles the *p*-orbital spin-density isosurfaces of electrons in *d*-type solid-state altermagnet candidates. The scalability of photonic crystals allows all geometric parameters to be expressed in terms of their ratio between the lattice constant. (b) Calculated helicity-polarized bands showing broken Kramers' degeneracy at $\Gamma M(M')$ intervals. The inset shows the layout of the simulated unit cell. (c) Isofrequency contours each sampled at $\omega a_0/2\pi c = 0.62$ (top) and $\omega a_0/2\pi c = 0.24$ (bottom). The symmetry-protected degeneracy is shown as four nodal points for each contour map.

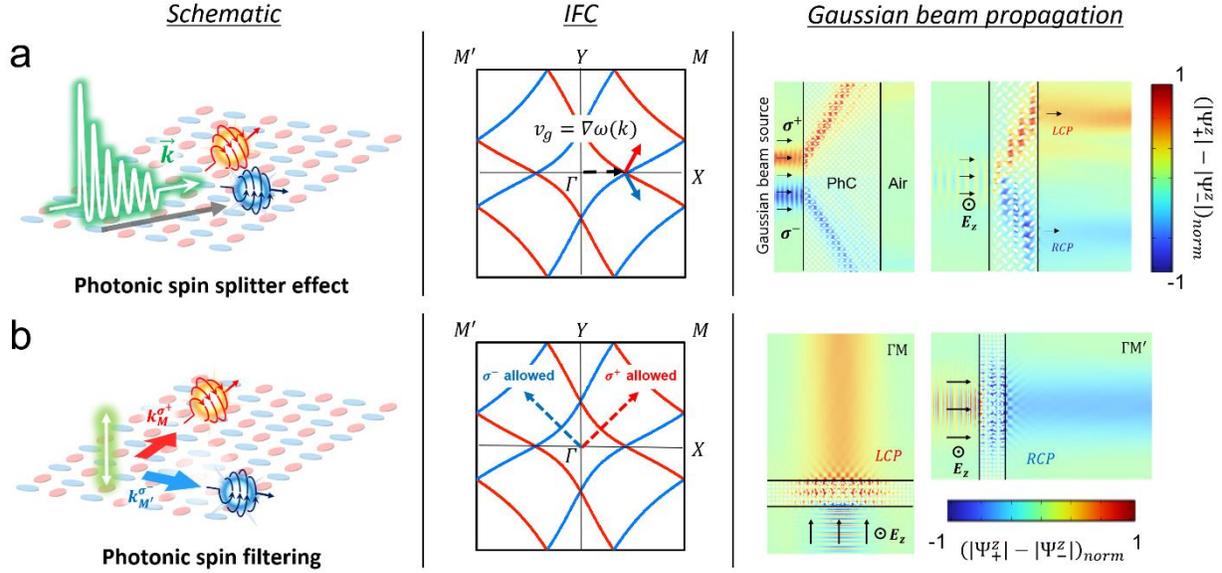

**Fig. 3.** Helicity- and momentum-dependent transport phenomena in photonic altermagnets. (a) Graphical depiction of the photonic spin-splitter effect (left). Helicity-dependent propagation directions are dictated by the IFC-dependent group velocities $v_g = \nabla\omega(k)$ (middle). Numerical simulations on helicity-polarized Gaussian beam propagation confirm that the propagation direction is normal to the spin-dependent IFC (right). Linearly polarized Gaussian beam therefore splits into two branches with opposite helicities. (b) Schematic of photonic spin filtering (left). One of the two helicity states is only allowed to propagate in the $\Gamma M(M')$ direction (middle), filtering out the orthogonal state through helicity-preserved reflection. Simulations show that a circularly polarized Gaussian beam exits the PhC for an incident linearly polarized light, where the helicity of the output beam changes according to the propagation direction. All simulation setups assume PhCs surrounded by air, as labelled on the simulation plots. The normalized spatial profile of the RS vectors color-coded by their corresponding helicities is plotted, where fields within the chiral objects are not plotted for better visual representation.